\documentclass[]{aanew}
\usepackage{psfig,graphicx}
\usepackage[authoryear]{natbib}
\bibpunct{(}{)}{;}{a}{}{,}
\begin{document}

\title{The Be/X-ray transient \object{KS 1947+300}\thanks{Partly
based on observations made with the Nordic Optical 
Telescope, operated on the island of La Palma jointly by Denmark,
Finland, Iceland, Norway, and Sweden, in the Spanish Observatorio del 
Roque de los Muchachos of the Instituto de Astrof\'{\i}sica de
Canarias.}} 

\author{I.~Negueruela\inst{1,2}
\and G.~L.~Israel\inst{3}
\and A.~Marco\inst{2}
\and A.~J.~Norton\inst{4}
\and R.~Speziali\inst{3}
}                   
                                                            
\institute{Observatoire de Strasbourg, 11 rue de l'Universit\'{e},
F67000 Strasbourg, France
\and
Dpto. de F\'{\i}sica, Ingenier\'{\i}a de Sistemas y 
Teor\'{\i}a de la Se\~{n}al, Universidad de Alicante, Apdo. de Correos 99,
E03080, Alicante, Spain
\and
Osservatorio Astronomico di Roma, Via Frascati 33, 
I00040 Monteporzio Catone, Italy
\and
Department of Physics \& Astronomy, The Open University, Walton Hall, 
Milton Keynes MK7 6AA, U.K.
}

\offprints{I. Negueruela,
\email{ignacio@disc.ua.es}}

\date{Received    / Accepted     }

\titlerunning{KS\,1947+300}
\authorrunning{Negueruela et al.}

\abstract{
We present optical spectroscopy and optical and infrared photometry
of the counterpart to the transient X-ray source 
\object{KS\,1947+300}. The counterpart is shown to be a moderately
reddened $V=14.2$ early-type Be star located in an area of low 
interstellar absorption slightly above the Galactic plane.
Changes in brightness are accompanied by correlated reddening of the
source, as is expected in this kind of object. From intermediate
resolution spectroscopy, we derive a spectral type B0Ve. If the
intrinsic luminosity of the
star is normal for its spectral type, \object{KS\,1947+300} is
situated at a distance of $\sim 10$ kpc, implying that its X-ray
luminosity at the peak of the spring 2000 X-ray outburst was 
typical of Type~II outbursts in Be/X-ray transients. 
\object{KS\,1947+300} is thus the first
Be/X-ray recurrent transient showing Type II outbursts which has an
almost circular orbit.
}
\maketitle 


\section{Introduction}

The transient hard X-ray source \object{KS\,1947+300} was discovered on
8th June 1989 by the TTM coded-mask X-ray spectrometer aboard the
{\em Kvant} module of the {\em Mir} orbiting space station \citep{bor90}.
The source was detected at a flux of $70\pm10\:{\rm mCrab}$ in the
$2-27$ keV range. Its spectrum could be approximated by a power law
with $\alpha=-1.72\pm0.31$ absorbed by a hydrogen column density
$N_{\rm H} =(3.4\pm3.0) \times10^{22}\:{\rm cm}^{-2}$ \citep{bor90}. The
source was detected in three further pointings of
the area during June and July 1989, but was not detected in August, 
when the 3-$\sigma$ upper limit on the flux was equivalent to 1/7th of
the flux observed in June \citep{bor90}.

\defcitealias{gor91}{GELS}
\defcitealias{gra91}{GSY}

The 30\arcsec\ error circle contains only two bright stars, which were
observed by \citeauthor{gra91} (\citeyear{gra91}, henceforth
\citetalias{gra91}). The brightest object was reported to have the
colours of a reddened distant early-type star. Independently,
\citeauthor{gor91} (\citeyear{gor91}, henceforth \citetalias{gor91})
searched a larger circle ($r=1\arcmin$)  
around the position of the X-ray source. They also concluded that the
same star (their Star 2) was the likely optical counterpart. A
spectrum of this object showed evidence for H$\alpha$ emission above
the night-sky level \citepalias{gor91}.

The transient X-ray pulsar \object{GRO J1948+32} was detected by the BATSE
detectors on board the {\em ComptonGRO} satellite on 6th April 1994
\citep{deepto}. The source displayed pulsations at 18.7 s and a hard
spectrum extending up to 75 keV. It was detected during 33 days
reaching a maximum flux of $\sim 50\:{\rm mCrab}$ in the $20-75$ keV
range. \citet{deepto} observed modulation of the neutron
star's pulse frequency suggestive of orbital variation. The very large
error box of \object{GRO J1948+32} (which included the TTM error
circle for \object{KS\,1947+300}) rendered the search for an optical
counterpart unfeasible, but the overall X-ray behaviour was
reminiscent of a Be/X-ray binary.

\begin{table*}
\caption{Photometric values for the optical counterpart to
  \object{KS\,1947+300} derived from our observations. 
  Uncertainties depend fundamentally on the
        calibration of the zero points and are upper limits.}
  \begin{center}
\begin{tabular}{lccccc}
Date& $U$ & $B$   &  $V$ &   $R$ &   $I$ \\
\hline
&&&&&\\
29/05/01&$14.98\pm0.05$&$15.17\pm0.05$&$14.24\pm0.08$& $13.53\pm0.10$& $12.89\pm0.05$\\
03/07/01&$14.76\pm0.08$&$15.06\pm0.05$&$14.16\pm0.03$&$13.50\pm0.02$&$12.77\pm0.04$
\end{tabular}
\end{center}
 \label{tab:phot}
\end{table*}

A new outburst of \object{KS\,1947+300} was detected by the All Sky
Monitor (ASM) on board the {\em RossiXTE} satellite starting around
23rd October 2000 \citep{lc00}. Further {\em RossiXTE} observations
revealed pulsations at $P_{\rm s} = 18.76\:{\rm s}$ \citep{sm00},
making the identity of \object{KS\,1947+300} and \object{GRO J1948+32}
virtually certain.  The outburst finished in late November 2000, but it
was shortly followed by a much larger one. Analysis of PCA/{\em RXTE}
pointed observations taken during this phase of activity has
resulted in the derivation of an orbital solution with $P_{{\rm
orb}}=40.43\:{\rm d}$ and $e<0.04$ \citep{gal}. 

In this paper we report on
optical and infrared observations of the proposed counterpart taken
during and after the outbursts. Our observations confirm the optical
counterpart and identify the system as a Be/X-ray transient.

\section{Observations}

\subsection{Optical Photometry}

Observations of the field were taken in service mode on the night of
29th May 2001 using the 1.0-m Jakobus Kapteyn Telescope (JKT) in La Palma,
Spain. We obtained images through $UBVRI$ filters with the JAG-CCD imaging
instrument equipped with a $2148 \times 2148$ SITe2 CCD.  As well as
individual frames covering the field of \object{KS\,1947+300}, 
we also observed three Landolt standard fields, containing a total of
nine standard stars \citep{landolt}, at a range of airmasses. 
Bias subtraction and flat fielding were carried
out on all frames using {\em Starlink} {\sc ccdpack} software
\citep{draper}. Then,
using the {\em Starlink} {\sc gaia} software \citep{draper2}, aperture
photometry was 
performed on all frames with background subtraction from annular sky
regions around each star.

Instrumental minus catalogue magnitudes were calculated for each of the
Landolt standard stars and linear fits were performed against airmass. The
resulting extinction coefficient and zero point corrections were
applied to the target star, resulting in the magnitudes shown in
Table~\ref{tab:phot}.  
Because the target frames were observed near the zenith,
uncertainties in the extinction coefficients have negligible effect,
and the uncertainties shown in the table are essentially due to zero
point uncertainties only. 

Observations of the field were also taken on the night of July 3rd 2001
using the 2.6-m Nordic Optical Telescope (NOT) in La Palma, Spain. We
obtained images through $UBVRI$ Bessel filters with the 
Andalucia Faint Object Spectrograph and Camera (ALFOSC), equipped with
a thinned $2048\times2048$ pixel Loral/Lesser CCD, covering a field of
view of $6\farcm4\times6\farcm4$.

\begin{figure*}[t]
\begin{picture}(500,260)
\put(0,0){\includegraphics{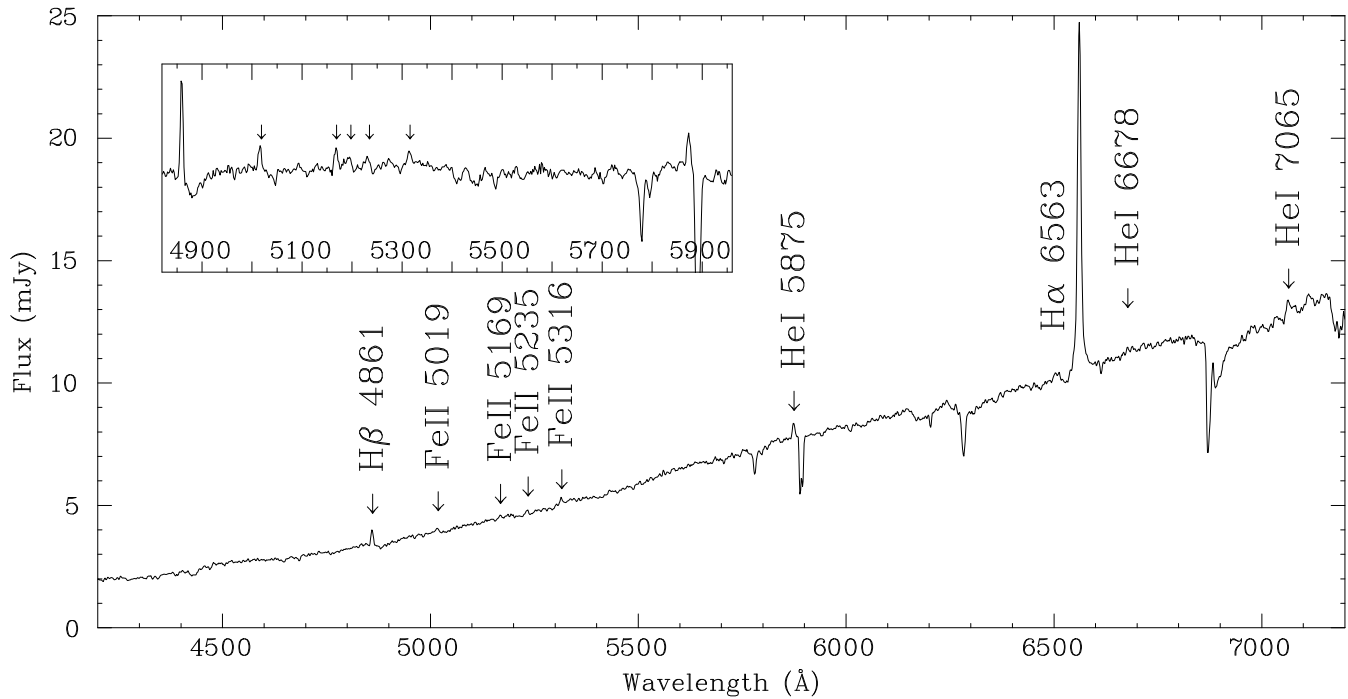}}
\end{picture}
  \caption{Flux-calibrated low-resolution spectrum of the optical
   counterpart to \object{KS\,1947+300}, obtained on February 8th,
   2001, using the INT+IDS. The calibration is not absolute due to
   slit losses. The inset shows in detail the emission lines in the
   yellow region, from a spectrum taken from Loiano on June 22nd,
   2001, using BFOSC + grating \#7. The positions of the \ion{Fe}{ii}
   blends at $\lambda\lambda$~5019, 5169, 5198, 5235 \& 5316\AA\ are
   indicated by arrows. } 
   \label{fig:lowres}
\end{figure*}

As well as individual frames covering the field of \object{KS\,1947+300}, 
we also observed three Landolt standard fields, each containing four
standard stars \citep{landolt}, at a range of airmasses. Photometric
reduction was carried out following 
an identical procedure to that described above for the JKT data. Similar
comments regarding the zero point uncertainties apply.

\subsection{Infrared Photometry}

The field of \object{KS 1947+300} was observed in the $J$, $H$ and $K$ 
filters (all with 60-s exposures) on 6th December 2000 from the 1.1-m 
AZT--24 telescope at Campo
Imperatore (Italy) equipped with the Supernova Watchdogging IR Camera
(SWIRCAM), which has a $4\farcm4\times4\farcm4$ field of view and
$1\farcs04$/pixel spatial resolution). 10\arcsec\ dithered
images were taken in the three filters. Data analysis procedures
similar to those described above were applied. A photometric standard 
was also observed (SAO 48300; 10 s in each filter) and used to derive
absolute magnitudes.
Within the X-ray positional uncertainty circle, we detected only one
bright ($H=11.4$) IR object, corresponding to the proposed optical
counterpart to \object{KS 1947+300}, which is thus the only reddened
star in the field. Its magnitudes are listed in Table~\ref{tab:ir}.

\begin{table}
\caption{Infrared photometric values for the optical counterpart to
  \object{KS\,1947+300} derived from our observations at Campo
        Imperatore.}
\begin{center}
\begin{tabular}{ccc}
$J$&$H$&$K$ \\
\hline
&&\\
$11.67\pm0.15$&$11.43\pm0.12$&$11.21\pm0.12$\\
\end{tabular}
\end{center}
 \label{tab:ir}
\end{table}

\subsection{Spectroscopy}

We obtained a low resolution spectrum of the optical counterpart on
December 1st 2000, using the 1.82-m telescope
operated by the Osservatorio Astronomico di Padova atop of Mount Ekar,
Asiago (Italy). The telescope was equipped with the Asiago Faint
Object Spectrograph and Camera (AFOSC) and the SiTE thinned CCD. We
used grism \#4 which gives a resolution of $\approx 8.3$\AA\ over the
$\lambda\lambda$ 3500\,--\,7500 \AA\ range. Details of this spectrum
were reported in \citet{iauc}.

A second spectrum (displayed in Fig.~\ref{fig:lowres})
was obtained on February 8th 2001 using the 2.5-m Isaac
Newton Telescope (INT), located  at the Observatorio del Roque de los 
Muchachos, La Palma, Spain. The telescope was equipped with the
Intermediate Dispersion Spectrograph (IDS)  with the 235-mm
camera. The choice of the R400V grating and thinned EEV\#10 CCD
results in a nominal dispersion of $\sim 1.4$ \AA/pixel. Measurements
of arc line widths indicate a spectral resolution of $\approx 5 $\AA\
(FWHM) at $\lambda$5500\AA. 

Further spectra were obtained on June 19-22, 2001, using the 1.52-m
G.~D.~Cassini telescope at the Loiano Observatory (Italy). The
telescope was equipped with the Bologna Faint Object Spectrograph and
Camera (BFOSC) and the new EEV camera. Several grisms were used,
giving different coverages and resolutions. A detail of the
higher resolution spectrum is shown in Fig.~\ref{fig:lowres}.

Finally spectra were obtained on the nights of July 2nd 2001 and
December 5-8th 2001 using ALFOSC on the NOT, equipped with grism
\#7. On the nights of July 2nd and December 5th, we used a $1\farcs8$
slit width, while on December 6th and 7th the slit width was
$1\farcs0$. The resolutions achieved with these configurations are
10.6\AA\ and 6.6\AA\ respectively.
Spectra taken with the $1\farcs0$ slit have been normalised and summed
and their blue end is displayed in Fig.~\ref{fig:blue}.

All the spectroscopic data were reduced with the {\em Starlink}
packages {\sc ccdpack}  and {\sc figaro}
\citep{shortridge} and analysed using {\sc figaro} and {\sc dipso}
\citep{howarth}.

\section{Results}

\subsection{Previous photometric work}
\citetalias{gra91}   and   \citetalias{gor91} independently observed
the error box of \object{KS\,1947+300} and report several measurements
of the optical counterpart. \citetalias{gor91} obtained $UBV$
photometry of the  source. They give 13 data-points  covering a span of
15 days starting  in September 15th 1990, followed by  a single
data-point 18 days later. 
\citetalias{gra91} give 14 sets of $UBVR$ data-points. The first eight
cover a span of 38 days starting on October 16th 1990, while the last
six cover 11 days separated from the first set by a 200-d gap  (i.e.,
in June 1991). The last point in the dataset of \citetalias{gor91}
overlaps in  time with \citetalias{gra91}'s first run. The values 
measured  by the two  teams on  JD 2448182  are compatible  within the
errors  quoted, showing that  their photometric  systems are  not very
different.

\begin{figure}[t]
\begin{picture}(250,250)
\put(0,0){\includegraphics{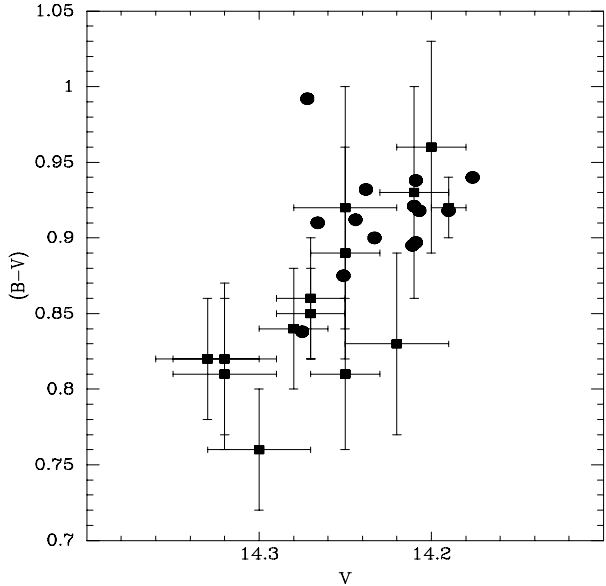}}
\end{picture}
\caption{A colour/magnitude plot of the photometric data presented by
\citetalias{gor91} and \citetalias{gra91}. The data from
\citetalias{gor91} are shown as filled squares with their associated
errors. The data from \citetalias{gra91}, who do not quote errors, are
shown as filled circles.
A tight correlation between the $V$ magnitude and the associated
reddening is obvious to the eye.} 
\label{fig:stern}
\end{figure}

Both sets of authors indicate that the dispersion in their
measurements is compatible with observational errors, while
\citetalias{gor91} explicitly mention that they do not observe
any clear correlation between the $B$
and $V$ bands. Close inspection of the datasets, however, shows that
both sets of authors clearly underestimated the accuracy of their
measurements. As can be seen in Fig.~\ref{fig:stern}, measurements
clearly display a very good correlation between $V$ and $(B-V)$. 
\citetalias{gra91} do not quote any errors for their photometry, but
the dispersion of data-points in 
Fig.~\ref{fig:stern} shows that they are unlikely to be any larger
than the errors in \citetalias{gor91}'s photometry. A
simple linear fit to all the data-points from both papers 
(without consideration of errors) gives a Spearson
correlation coefficient $R=0.64$, clearly suggesting that there is an
underlying correlation. 

\begin{figure}[t]
\begin{picture}(250,250)
\put(0,0){\includegraphics{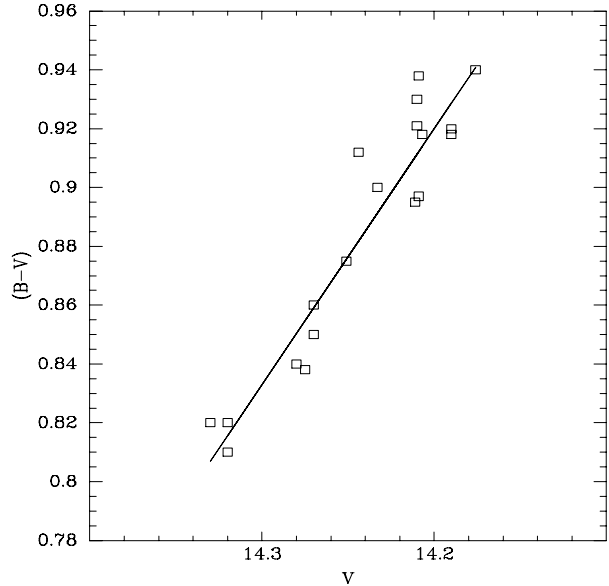}}
\end{picture}
\caption{Colour/magnitude plot of the combined photometric datasets of
\citetalias{gor91} and \citetalias{gra91}. The best linear fit,
corresponding   to 
Eq.~\ref{eq:rusos}, is given as a straight line. The Spearson
correlation coefficient for the fit is $R=0.95$, indicating that there
is a very tight correlation between $V$ and $(B-V)$.}
\label{fig:clean}
\end{figure}

Even though the errors of some measurements are relatively
large, none of the values deviates strongly from the
general trend. As a consequence, it was thought appropriate to improve
the quality of the fit by following an iterative procedure: a 
linear fit to all the points was obtained, the data-point that
deviated most strongly from the fit was removed and the
remaining points were fitted, repeating
until no further improvement in the correlation coefficient was
obtained. Only a few points had to be removed before the
correlation coefficient was considerably improved. For example,
removal of a single deviating point seen at the top of
Fig.~\ref{fig:stern} improved the correlation coefficient from
$R=0.64$ to $R=0.77$. 
Figure~\ref{fig:clean} shows the final linear fit to 19 
data-points (taken approximately evenly from both sets). The goodness
of the fit is indicated by a Spearson correlation coefficient
$R=0.95$. Interestingly, the best fit line, given by 
\begin{equation}
(B-V)= 13.3-0.87V
\label{eq:rusos}
\end{equation}
is identical in slant and zero point to the original fit to the
complete dataset, clearly showing that our selective analysis has
not deleted any information contained in the plot. Such a result
suggests that any dispersion in the photometry due to
instrumental errors has not been systematic and the few deviating
points may be attributed to low photometric quality.

From the strength of the correlation and the fact that the coefficient
multiplying $V$ is rather close to unity, it can be deduced that
most of the variation in $(B-V)$ is simply due to variability in $V$
and that the variation in $B$ is very small in comparison. The whole
behaviour is consistent with what is expected from a Be star, where
the variability is due to emission from a circumstellar disk with a
temperature $T_{{\rm disk}} \approx 0.5 T_{{\rm eff}}$, where $T_{{\rm
    eff}}$ is the effective temperature of the star. The disk
contribution to the total spectral energy distribution is very small
in the $B$ band and increases towards longer wavelengths. The 
spectral energy distribution (star plus disk)
becomes bluer when it is fainter because the contribution from the
disk decreases.

For this reason, rather than taking an average of all the photometric
measurements of the star, we consider that the faintest (and bluest)
points in the dataset are more representative of the intrinsic
magnitudes of the Be star, since the contribution from the disk is
smallest. The three points in the bottom left region of
Figure~\ref{fig:clean} average to $V=14.32\pm0.02$
and $(B-V)=0.82\pm0.02$, which is compatible within the errors with
the linear relationship found above.  

\subsection{Spectral classification}
An intermediate resolution spectrum of the optical counterpart to
\object{KS\,1947+300} is displayed in Fig.~\ref{fig:lowres}. Except
for some variability in the strength of the H$\alpha$ line (see
Section~\ref{sec:halpha}), all our spectra are nearly identical.
The characteristics are typical of an early-type reddened distant Be
star: H$\alpha$ and H$\beta$ appear strongly in emission, as
also does
\ion{He}{i}~$\lambda$5875\AA. \ion{He}{i}~$\lambda$7065\AA\ and
\ion{He}{i} $\lambda$6678 \AA\ can be seen as very weak emission
features in the spectra with highest SNR. Some \ion{Fe}{ii} lines
(indicated in Figure~\ref{fig:lowres}) can be seen in emission, as
well. The continuum is characterised by strong diffuse
interstellar bands. The presence of \ion{He}{i} emission implies a
spectral type earlier than B2, putting the source in the spectral
range of counterparts to Be/X-ray binaries \citep{yo02}.

\begin{figure*}
\begin{picture}(500,260)
\put(0,0){\includegraphics{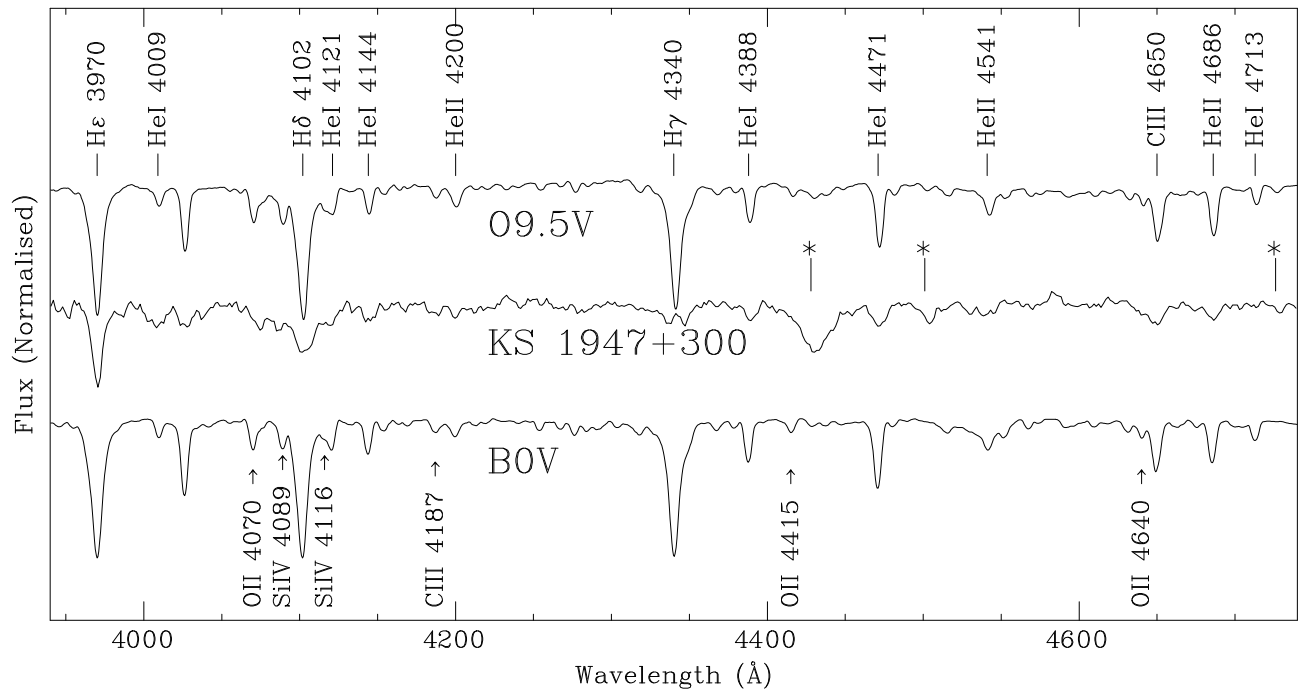}}
\end{picture}
  \caption{Blue spectrum of the optical counterpart to
   \object{KS\,1947+300} compared to those of the B0V standard and the
   O9.5V standard artificially spun up to $v\sin i =220\:{\rm
   km}\,{\rm s}^{-1}$ in order to reproduce the very broad and shallow
   absorption lines in the spectrum of \object{KS\,1947+300} (as is
   typical in  Be stars). Diffuse Interstellar Bands
   are marked with a `*'.}
   \label{fig:blue}
\end{figure*}

Absorption lines typical of an early-type star can be seen in the blue
part of the spectrum. A higher resolution spectrum of this region is
displayed in Fig.~\ref{fig:blue}. The presence of \ion{He}{ii} implies
a spectral type earlier than B0.5. 
 The ratio between \ion{He}{ii}~$\lambda$4686\AA\ and
\ion{C}{iii}~$\lambda$4650\AA\ is only compatible with a spectral type
B0 for a main-sequence object. At O9.5V, \ion{He}{ii}~$\lambda$4686\AA\
is already stronger than the \ion{C}{iii} line \citep{waf}, while the
presence of \ion{He}{ii}~$\lambda$4200\AA\ is incompatible with a
spectral type later than B0.2. Though the strength of the
\ion{Si}{iv} doublet on the wings of H$\delta$ could suggest an
O9.5III spectral class, the ratio between
\ion{He}{ii}~$\lambda$4200\AA\ and \ion{C}{iii}~$\lambda$4187\AA\ and
the weakness of \ion{He}{ii}~$\lambda$4541\AA\ support the earlier
spectral type.
Therefore we adopt a B0Ve spectral type.

\subsection{Evolution of H$\alpha$}
\label{sec:halpha}

Profile changes in emission lines, particularly H$\alpha$, may be used
to trace the dynamical evolution of the Be envelope
\citep[e.g.,][]{yo01}. Unfortunately, the resolution of most of our
spectra is too low for this aim, but some information may still be
collected from the evolution of the line strength.

\begin{table}
\caption{EW of H$\alpha$ measured on our spectra. Errors are estimated
from the spread of values obtained by using different measurement
methods and selecting different continua.}
  \begin{center}
\begin{tabular}{lc}
Date& EW (\AA)\\
\hline
&\\
Dec. 1, 2000&$-14.7\pm1.0$\\
Feb. 8, 2001 &$-14.8\pm0.5$\\
Jun 19, 2001 & $-15.1\pm1.0$\\
Jun. 22, 2001 &$-15.5\pm0.8$\\
Jul. 7, 2001 & $-16.5\pm0.5$\\
Dec. 7, 2001 & $-15.3\pm0.5$
\end{tabular}
\end{center}
 \label{tab:ha}
\end{table}

Table~\ref{tab:ha} displays the Equivalent Width (EW) of H$\alpha$ for
our spectra. Though the data are sparse, it seems likely that the
strength of the line must have remained relatively constant for the
whole period, with a slight increase around July 2001 (i.e., after the
end of the large X-ray outburst).

\section{Discussion}

\object{KS\,1947+300} is a transient X-ray source, which has appeared
three times at very high luminosity and is at most weakly
detected (below 6 mCrab) by the {\em RXTE}/ASM during quiescence
\citep{lc00}. This behaviour is typical of the class of 
Be/X-ray transients, like \object{4U\,0115+63} or \object{A\,0535+26}
(see \citealt{we01}). \object{KS\,1947+300} has only been observed on
three occasions, in 
1989, 1994 and 2000-2001. Such a recurrence timescale is typical
of Type~II (or giant) outbursts in Be/X-ray transients.

\begin{figure}
\begin{picture}(250,175)
\put(0,0){\includegraphics{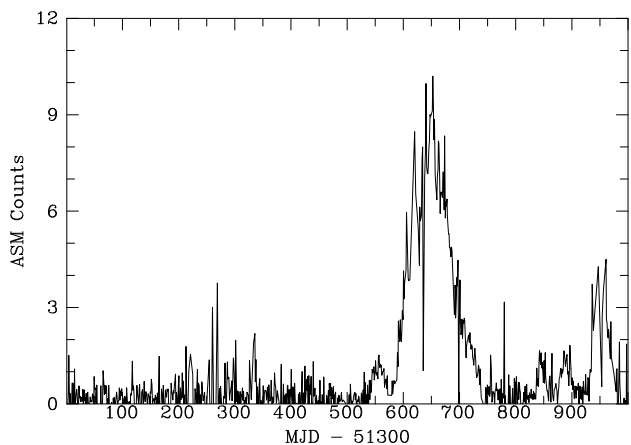}}
\end{picture}
  \caption{
{\em RXTE}/ASM X-ray lightcurve of 
   \object{KS\,1947+300} between MJD 51300 (May 2nd, 1999) and
 MJD 52300 (Jan 26th 2002). Data are one-day averages of the ASM three
energy-bands. Two major outbursts, centred on February 13th 2001 and
December 17th 2001, as well as several minor flares are visible after
MJD 51800.}
   \label{fig:asm}
\end{figure}

\object{KS\,1947+300} became active in late 2000. After a first weak
outburst in November 2000, it underwent a very long giant outburst,
lasting close to 150 days and reaching a flux in excess of $120\:{\rm
mCrab}$ around MJD 51953 (February 13th). Two further weak outbursts and a
moderate-intensity one followed during the second half of 2001 (see
Fig.~\ref{fig:asm}).  

Our observations have shown that the optical counterpart to
\object{KS\,1947+300} is a moderately reddened B0Ve star.
For the adopted spectral type, an intrinsic colour $(B-V)_{0}=-0.27$
may be considered \citep{weg94}. We can then assume that the observed
faintest photometric dataset corresponds to the intrinsic magnitudes
and colours of the star (obviously, if this is not the case, the star
will be intrinsically fainter and bluer, allowing the distance
derived here to be used as a lower limit). Therefore we derive an excess
$E(B-V)=1.09$. The ratio of selective to total extinction $R_{V}$ in
this area has been shown to be close to standard out to moderate
distances \citep{tur76}. Hence we will use the 
standard law (i.e., $R=3.1$ -- see \citealt{fit99}), resulting in
$A_{V}=3.38$.  With an average absolute magnitude for B0V $M_{V}=-4.2$
\citep{vac96}, we derive then a 
distance of $\approx 10$ kpc to \object{KS\,1947+300}.

With galactic coordinates ($l=66\degr.1$, $b=+2\degr.1$),
\object{KS\,1947+300}  
lies in an area of rather low obscuration. Extinction is moderate
within one kpc of the Sun and then remains constant out to at least 5
kpc \citep{for85}, which explains why such a distant object is
moderately bright. In this area of the sky, the Perseus Arm is defined
by the Vul OB2 association, which, with a dereddened distance modulus
$DM=13.2$, lies at 4.4 kpc \citep{tur80}. The average reddening of six
Vul OB2 stars lying at $b>1\degr$ is $E(B-V)=0.6$ \citep{tur80}. The
rather larger reddening to \object{KS 1947+300} supports a higher
distance. The source is then likely to be located in the Cygnus or
Outer Arm, which
in this direction is located at $\approx9$ kpc \citep{tac93}, in good
agreement with our distance estimate.
The distance estimated for \object{KS\,1947+300} is very similar to
that found for the nearby source \object{XTE J1946+274} \citep{franz},
which should also lie on the same arm. 

Considering this distance, the peak X-ray luminosity (uncorrected
for effects of circumstellar or interstellar absorption) observed
by {\em BeppoSAX} during the March 2001 outburst was $L_{\rm x}
\approx 2\times10^{37}\:{\rm erg}\,{\rm s}^{-1}$ (Rea et al., in
prep.), which is a typical value for Be/X-ray transients 
during a Type II outburst.  The occurrence of
weaker outbursts after a Type II outburst is also a typical feature of a
subgroup of Be/X-ray transients, characterised by close orbits with
moderate and low eccentricity \citep{we01}. The only peculiarity of
this outburst was its duration, rather longer than is typical in other
Be/X-ray transients (typically 4$-$6 weeks).

The occurrence of several X-ray outbursts has not been reflected in
any changes in the intensity of the H$\alpha$ line during the period
considered. Though obvious correlations have been seen in other
systems (e.g., \object{4U 0115+63}; \citealt{yo01}), such effects are
likely to be strongly dependent on the relative geometry of the
system: if the Be star is seen
under a relatively low inclination angle, any changes in the Be disk
structure would be barely reflected in the measured properties of
H$\alpha$. It is a well known fact, as derived from both
observations \citep{hanal96} and modelling \citep{hh97}, that at
high and moderately high inclinations optically thick lines (such as
H$\alpha$) show the effects of density perturbations only as
relatively weak flank inflections, as the range of projected
rotational velocities is small and several radiative transfer effects
(such as non-coherent scattering; c.f. \citealt{hh97}) result in the
broadening of features. Such inflections will certainly not be
observable at the moderate resolutions used here. Higher resolution
spectroscopy will be therefore necessary in order to derive a value
for the counterpart's $v\sin i$.

The main difference between \object{KS\,1947+300} and other Be/X-ray
transients with similar behaviour lies on its very low orbital
eccentricity $e<0.04$ \citep{gal}. Few Be/X-ray transients
with low eccentricities are known. An upper limit $e\la 0.09$ was set
by \citet{kel83} for the orbit of \object{2S 1553$-$542}, an X-ray
transient observed in 1975 during a single outburst, which was likely
to be a Be/X-ray binary. The 27.1-s X-ray pulsar \object{XTE
J1543-568} is also likely a 
Be/X-ray binary. It displayed a large outburst in 2000, followed
by weaker activity \citep{zand01}. \object{XTE J1543-568} has a 75.6-d
orbital period with an eccentricity
$e<0.03$. As both \object{2S 1553$-$542} and \object{XTE
J1543-568} have only been observed during one giant outburst each,
\object{KS\,1947+300} is the first
low-eccentricity recurrent transient displaying Type~II outbursts.

\citet{we01} argued that one could debate whether the preponderance of
moderately eccentric orbits among Be/X-ray transients was an
observational effect or reflected the actual distribution. Since the
neutron star companion is very effective at truncating the
circumstellar disk of the Be star when the eccentricity is low, it
could well be that low-eccentricity systems rarely display bright
X-ray outbursts and are difficult to detect. The behaviour of
\object{KS\,1947+300} seems to argue otherwise, since this system with
a practically circular orbit has displayed 3 Type II outbursts in 11
years, a recurrence timescale comparable to those of the most active
Be/X-ray transients. 

The implication is then that low-eccentricity systems are as likely to
display Type II outbursts as systems with moderate eccentricity and
therefore the small number of low-eccentricity systems detected is
actually reflecting the dominance of moderately eccentric orbits among
Be/X-ray transients. This would then be one further argument in favour
of supernova kicks \citep[c.f.,][]{vv97,yo02}.

\section{Conclusions}
We have shown that the optical counterpart to the recurrent X-ray
transient \object{KS\,1947+300} is a B0Ve star at an approximate distance
of 10 kpc. \object{KS\,1947+300} is therefore a Be/X-ray transient
displaying recurrent Type~II outbursts with intrinsic luminosities
similar to other bright Be/X-ray transients. Among the $\sim 10$
systems known displaying this kind of behaviour, \object{KS\,1947+300}
is the first one to have an almost circular orbit. Since its existence
shows that Be/X-ray binaries with very low eccentricities can be
detected as bright X-ray sources, it is to be inferred that the
preponderance of systems with moderate eccentricities among the
observed sample reflects the actual distribution of orbital
eccentricities.

\begin{acknowledgements}

Part of the data presented here have been taken using ALFOSC, which is 
owned by the Instituto de Astrof\'{\i}sica de Andaluc\'{\i}a (IAA) and
operated at the Nordic Optical Telescope under agreement
between IAA and the NBIfAFG of the Astronomical Observatory of
Copenhagen. 
The INT and JKT are operated on the island of La 
Palma by the Isaac Newton Group in the Spanish Observatorio
del Roque de Los Muchachos of the Instituto de
Astrof\'{\i}sica de Canarias.
The G.D. Cassini telescope
is operated at the Loiano Observatory by the Osservatorio Astronomico di
Bologna.

The X-ray results were provided by the ASM/RXTE teams at MIT and at
the RXTE SOF and GOF at NASA's Goddard Space Flight Center through the
High Energy Astrophysics Science Archive Research Center Online
Service, which is made available by the NASA/Goddard Space Flight
Center.  

 This research has made use of the 
Simbad data base, operated at CDS,
Strasbourg, France. 

 We thank the referee, Dr. A.~M.~Levine, for suggestions that helped
to improve the paper.

\end{acknowledgements}

\end{document}